\documentclass{article}
\usepackage{spconf,amsmath,graphicx}
\usepackage{lipsum}  
\usepackage[utf8]{inputenc}
\usepackage{cite}
\usepackage{graphicx}
\usepackage{textcomp}
\usepackage{url}
\usepackage{epsfig}
\usepackage{float}
\usepackage{xcolor}
\usepackage{balance}
\usepackage{bbm}
\usepackage{textcomp}
\usepackage{enumitem}
\usepackage{mathtools}
\usepackage{algorithm,algcompatible,amssymb,amsmath}
\usepackage{upgreek}

\algnewcommand\algorithmicto{\textbf{to}}
\algnewcommand\RETURN{\State \textbf{return} }

\title{Lyapunov-driven deep reinforcement learning for edge inference empowered by Reconfigurable Intelligent Surfaces}
\name{Kyriakos Stylianopoulos, Mattia Merluzzi, Paolo Di Lorenzo, and George C. Alexandropoulos\thanks{
	 K.~Stylianopoulos and G.~C.~Alexandropoulos are with the Department of Informatics and Telecommunications, National and Kapodistrian University of Athens, 15784 Athens, Greece (e-mail: kstylianop; alexandg\}@di.uoa.gr).
	 G.~C.~Alexandropoulos is also with the Technology Innovation Institute, Abu Dhabi, UAE.
	 M. Merluzzi is with CEA-Leti, Universite Grenoble Alpes, F-38000 Grenoble, France (email: mattia.merluzzi@cea.fr).
	 P. Di Lorenzo is with Consorzio Nazionale Interuniversitario per le Telecomunicazioni (CNIT), and Sapienza University of Rome, Italy (email: paolo.dilorenzo@uniroma1.it). This work was supported by the EU H2020 RISE-6G project under the grant number 101017011.
}}

\address{\vspace{-25mm}}

\begin{document}
\ninept
\maketitle
\begin{abstract}

In this paper, we propose a novel algorithm for energy-efficient, low-latency, accurate inference at the wireless edge, in the context of 6G networks endowed with reconfigurable intelligent surfaces (RISs). We consider a scenario where new data are continuously generated/collected by a set of devices and are handled through a dynamic queueing system. Building on the marriage between Lyapunov stochastic optimization and deep reinforcement learning (DRL), we devise a dynamic learning algorithm that jointly optimizes the data compression scheme, the allocation of radio resources (i.e., power, transmission precoding), the computation resources (i.e., CPU cycles), and the RIS reflectivity parameters (i.e., phase shifts), with the aim of performing energy-efficient edge classification with end-to-end (E2E) delay and inference accuracy constraints. The proposed strategy enables dynamic control of the system and of the wireless propagation environment, performing a low-complexity optimization on a per-slot basis while dealing with time-varying radio channels and task arrivals, whose statistics are unknown. Numerical results assess the performance of the proposed RIS-empowered edge inference strategy in terms of trade-off between energy, delay, and accuracy of a classification task. 
\end{abstract}

\vspace{-2mm}
\section{Introduction}
\label{sec:intro}
\vspace{-2mm}

The development of the next generation of wireless communication systems, known as 6G, is still at its infancy. The main challenge of 6G is to provide an Artificial Intelligence (AI) and Machine Learning (ML) native communication infrastructure. This concept is known as Edge ML/Edge AI \cite{Letaief22}. Edge AI comes with a twofold perspective, including both the benefits of AI/ML algorithms exploited for network optimization and orchestration, and the benefit of a powerful and efficient communication platform to process and distill large volumes of data collected by heterogeneous devices. The final aim is to accurately perform learning tasks within low end-to-end (E2E) delays, in the most efficient way from different perspectives entailing energy, communication overhead, etc. Edge AI will strongly benefit from Multi-access Edge Computing (MEC) \cite{ETSIMEC}, thanks to the deployment of distributed computing resources close to end users, namely in Mobile Edge Hosts (MEHs) that are, e.g. co-located with radio Access Points (APs). This will allows mobile end devices to access computing resources, albeit limited with respect to central clouds, in a fast, secure,  reliable, and possibly sustainable manner. In this work, we focus on both roles of edge AI in future systems, devising a resource allocation framework to enable dynamic energy efficient edge classification on data collected by end devices, with target E2E delays and inference reliability constraints. Besides the paradigm of communications for AI, in this work, the twofold role of edge AI is explored through the marriage between model-based Lyapunov stochastic network optimization, and Deep Reinforcement Learning (DRL), with the latter playing the role of an AI-based optimization algorithm, able to compensate the limitations of model-based optimization, namely complexity and/or lack of system modeling.
Previous works successfully attempted to merge such tools \cite{Bi2021,Sana2021}, however never focusing on the edge AI use cases, i.e. not taking into account application performance. Instead, the goal of this paper is to focus on the performance (in terms of accuracy) of an edge classification task, along with typical MEC performance including End-to-End (E2E) delay and energy consumption. 

\noindent\textbf{Related works.} As illustrated in the overview papers \cite{Zhou19_EI,Letaief22}, an efficient design of edge-inference hinges on several aspects, such as memory footprint optimization \cite{Han15}, adaptive model selection \cite{Tay18}, or goal-oriented optimization, for instance adapting frame rate and resolution of video streaming for efficient inference \cite{Jia18}, \cite{Ran18}. The authors of \cite{Gal21} consider a video analytics task, maximizing the average accuracy under a frame rate and delay constraint. 
Also, the work in \cite{MerluzziEML2021} proposed a joint management of radio and computation resources for edge ML, hinging on data quantization to control the accuracy of several learning tasks. Finally, the closest reference \cite{Merluzzi_ICC2022} proposes a dyanmic joint optimization or radio, computing resources, and JPEG data compression, fully based on Lyapunov optimization, assuming full knowledge of the system model.


All the aforementioned works enabled edge computing and/or inference considering the presence of a suitable wireless propagation environment. Moving toward millimeter wave communications (and beyond), the performance of MEC-oriented systems can be severely reduced due to poor channel conditions and blocking events. In this context, a strong performance improvement can be obtained exploiting Reconfigurable Intelligent Surfaces (RISs) \cite{Marco2019,WavePropTCCN,RISE-6G-EUCNC, RISE-6G-Commag,Tsinghua_RIS_Tutorial}, which are programmable surfaces made of hundreds of nearly passive reflective elements controlled to realize dynamic transformations of the wireless propagation environment, both in indoor and outdoor scenarios.
The inclusion of RISs in MEH systems offer a two-fold benefit: i) it alleviates the effect of blockages, which leads to low offloading rates and poor performance; ii) it enables better exploitation of the computing resources of the edge server thanks to the improved offloading capabilities.
Several works in the literature have already exploited RISs to empower wireless communications \cite{huang2019reconfigurable,wu2019intelligent,Pervasive,DCB}) and, very recently, also for computation offloading \cite{bai2020latency,di2021dynamic}. Finally, preliminary results on edge learning empowered by RISs appear in \cite{huang2021reconfigurable}, which considered static resource allocation for edge inference, and in the work \cite{battiloro2022lyapunov} that instead focused on adaptive federated learning.

\noindent\textbf{Contributions.} In this paper, we focus on an edge inference task, aimed at performing classification of data collected, distilled, and processed at the edge of a wireless network endowed with MEC capabilities and RISs. Our design aims at striking the minimum average power necessary by the system to perform low-latency inference with accuracy constraints. To this aim, we propose an online method based on the marriage of Lyapunov stochastic optimization DRL, to dynamically optimize: i) data compression scheme; ii) users’ transmission precoding and power; iii) RISs’ reflectivity parameters; iv) MEC resources. Our method leads to the convergence of model-based (i.e., Lyapunov) and data-driven (i.e., DRL) stochastic optimization into a single holistic framework, thus paving the way towards fully reconfigurable networks endowed with effective and efficient edge AI capabilities. The main novelties with respect to the closest work \cite{Merluzzi_ICC2022} are the following: i) the function linking accuracy and compression scheme is supposed to be unknown, and optimized through a DRL-based approach; ii) the RIS is not present in \cite{Merluzzi_ICC2022}, and it is optimized here through the DRL-based approach, differently from \cite{Airod2022}, in which it is optimized though a projected gradient descent method. The computation part is handled as in \cite{Merluzzi_ICC2022,Airod2022}, thus it does not represent a novelty in this paper.
Finally, numerical results assess the performance of the proposed method. \\
\textbf{Notation.} Bold upper case letters denote matrices, $\textrm{Tr}(\cdot)$ denote the trace operator, $|\cdot|$ denotes the determinant, and the superscript $(\cdot)^H$ denotes the hermitian operator; $\mathbf{I}_N$ and $\mathbf{0}_N$ denote the $N\times N$ identity matrix and the all-zeros matrix, respectively.
Finally, given a variable $X(t)$, $\overline{X}=\lim_{T\to\infty}\frac{1}{T}\sum_{t=1}^T\mathbb{E}\{X(t)\}$ if the sequence converges, or $\overline{X}=\lim \ {\rm sup}_{T\to\infty}\frac{1}{T}\sum_{t=1}^T\mathbb{E}\{X(t)\}$ otherwise.

\vspace{-2mm}
\section{System Model}\label{sec:system}
\vspace{-2mm}

Let us consider a wireless setup with a set $\mathcal{K}$ of Mobile Devices (MDs), each one equipped with $N_u$ antennas, an $N_a$ antennas Access Point (AP), with a co-located MEH able to process tasks through a pre-trained and pre-uploaded ML model (e.g. a deep neural network). Finally, an RIS with $M$ reflecting unit elements is assumed to be available, and it can be dynamically reconfigured to enhance MEC service performance. As in \cite{Merluzzi_ICC2022}, we model the inference process as a flow of request arrivals, which are first buffered locally at each MD before transmission and, after transmission, buffered remotely before computation. Time is organized in slots $t$ of equal duration $\tau$. Each device has its own buffer and all MDs compete for radio and computing resources. Then, uplink communication, along with computation, is the focus of this work.

\vspace{-2mm}
\subsection{Channel model}
\vspace{-2mm}

An RIS-aided wireless channel can be modeled by two components: i) a direct path from MD $k$ to the AP, whose time-varying coefficient are the elements of a matrix $\mathbf{H}_{k,d}(t)\in\mathbb{C}^{N_a\times N_u}$, and ii) an indirect path, created by the reflection of the RIS. The latter includes a channel matrix $\mathbf{H}_{k,r}(t)\in\mathbb{C}^{M\times N_u}$ from MD $k$ to the RIS, and a channel matrix $\mathbf{H}_{r,a}(t)\in\mathbb{C}^{N_a\times M}$ from the RIS to the AP. Finally, the RIS is composed of passive elements, whose phases can be dynamically and opportunistically reconfigured, and its response can be written as a diagonal matrix $\mathbf{\Phi}(t)$, with diagonal elements $\{\mathbf{\Phi}_{i,i}(t)=e^{j\phi_i(t)}\}_{i=1}^{M}$, where $\phi_i(t)$ denotes the (reconfigurable) phase shift of elements $i$. Then, the overall channel is \cite{Pervasive}
\begin{equation}\label{channel_model}
    \mathbf{H}_k(t)=\mathbf{H}_{k,d}(t)+\mathbf{H}_{r,a}(t)\mathbf{\Phi}(t)\mathbf{H}_{k,a}(t), \quad \forall k\in\mathcal{K}.
\end{equation}
We assume that, at each slot, every MD selects a transmit precoding strategy, based on current connect-compute system conditions. Then, denoting by $\mathbf{F}_k(t)$ the input covariance matrix of MD $k$, the instantaneous data rate can be written as follows:
\begin{equation}\label{data_rate}
    R_k(t)=W\log_2\left|\mathbf{I}_{N_u}+\frac{1}{\sigma^{2}}\mathbf{H}_k(t)\mathbf{F}_k(t)\mathbf{H}_k^H(t)\right|,\;\;\forall k\in\mathcal{K}
\end{equation}
where $\sigma^2=N_0W$, with $N_0$ the noise power spectral density, and $W$ bandwidth. All users are coupled by the RIS reflection.

\vspace{-2mm}
\subsection{Communication and computation queuing models}
\vspace{-2mm}

Similarly to \cite{Merluzzi_ICC2022}, buffers are intended as units of patterns (or data unit). At each slot, a generic MD $k$ accepts $A_k(t)$ new patterns into its communication buffer, while transmitting previously buffered patterns through the (RIS-aided) wireless connection with the AP, at a rate $R_k(t)$ (cf. \eqref{data_rate}). Then, denoting by $Q_k^l(t)$ the buffer size at time $t$, the communication queue evolves as follows:
\begin{equation}\label{local_queue}
    Q_k^l(t+1)=\max\left(0,Q_k^l(t)-\left\lfloor\frac{\tau R_k(t)}{n_{k,b}(c_k(t))}\right\rfloor\right)+A_k(t),
\end{equation}
where $n_{k,b}(c_k(t))$ is the number of bits encoding all patterns transmitted at time $t$, which is a function of the data compression level $c_k(t)\in\mathcal{C}_k$
at time slot $t$. The transmitted patterns join a remote computation queue, which is drained by the MEH's processing, i.e. the issuing of inference results. Then, denoting by $f_k(t)$ the MEH's CPU cycle frequency assigned to user $k$, the queue evolves as:
\begin{align}\label{remote_queue}
    Q_k^r(t+1)=&\,\max
\left(0,Q_k^r(t)-\left\lfloor\frac{\tau f_k}{w_k}\right\rfloor\right)\nonumber\\
&\quad+\min\left(Q_k^l(t),\left\lfloor\frac{\tau R_k(t)}{n_{k,b}(c_k(t))}\right\rfloor\right),
\end{align}
where $w_k$ is the computation load, i.e. the number of CPU cycles needed to output one inference result. 

\vspace{-2mm}
\subsection{Inference performance indicators}
\vspace{-2mm}

Effective and efficient edge inference entails timing (i.e., the delay from request issuing at the device until its treatment at the MEH), accuracy (e.g., correctly classified patterns), and energy. The average delay entails communication and computation phases. Also, if the queues are strongly stable, i.e. $\overline{Q}_k^{l(r)}<\infty$, the average E2E delay is finite and can be written in closed form thanks to Little's law: $\overline{D}_k=\tau(\overline{Q}_k^l+\overline{Q}_k^r)/\overline{A}_k$, where $\overline{A}_k$ denotes the average number of arrivals per slot. In this paper, we consider a generic metric $\mathcal{G}_k(c_k)$ of inference reliability, which we mildly assume to be a function of the employed compression scheme $c_k$. Such generic function could be the accuracy of a classification/regression/estimation task, or any other measure reflecting effective operation of an ML model running in the MEH (e.g., classification confidence). The aim will be to keep the long-term average of the accuracy metric $\overline{\mathcal{G}}_k(c_k)$ above a predefined threshold, set a priori as an application requirement. Finally, as source of MDs' power consumption we consider the long-term average transmit power, which can be expressed as $\overline{\sum_{k\in\mathcal{K}}\textrm{Tr}(\mathbf{F}_k)}$.

\vspace{-2mm}
\section{Problem formulation}\label{sec:prob_formulation}
\vspace{-2mm}

In this work, our aim is to guarantee energy-efficient edge inference with a minimum level of accuracy and a given E2E delay. The problem is formulated as follows:
\begin{align}\label{prob_form}
    &\hspace{-.5 cm}\underset{\{\mathbf{F}(t)\}_{k,t},\{f_k(t)\}_{k,t},\{\phi(t)\}_{i,t},\{c_k(t)\}_{k,t}}{\min}\;\;\overline{\sum_{k\in\mathcal{K}}{\textrm{Tr}(\mathbf{F}_k)}}\\
    &\textrm{subject to}\nonumber\\
    &(a)\;\overline{Q}_k^l<\infty,\;\forall k\in\mathcal{K}\qquad\qquad(b)\;\overline{Q}_k^r<\infty,\;\forall k\in\mathcal{K}\nonumber\\
    &(c)\;\overline{\mathcal
    G}_k(c_k)\geq \mathcal{G}_{k,\textrm{th}},\;\forall k\in\mathcal{K}\quad\;\; (d)\;\mathbf{F}_k(t)\succcurlyeq 0,\;\forall k\in\mathcal{K}\nonumber\\
    &(e)\;\textrm{Tr}(\mathbf{F}_k(t))\leq P_k,\;\forall k\in\mathcal{K} \quad(f)\;c_k\in\mathcal{C}_k,\;\forall k\in\mathcal{K}\nonumber\\
    &(g)\;\; \phi_i(t)\in[0,2\pi],\; i=1,\ldots M\nonumber \nonumber\\
    &(h)\;f_k(t)\geq 0\quad(i)\;\sum\nolimits_{k\in\mathcal{K}}f_k(t)\leq f_{\max}\nonumber
    \end{align}
Besides queues' stability in $(a)$ and $(b)$, the constraints of \eqref{prob_form} have the following meaning: $(c)$ the average accuracy is higher than a predefined threshold; $(d)$ the covariance matrix of each user is positive semidefinite; $(e)$ the instantaneous transmit power of each user is lower than a threshold; $(f)$ the compression scheme belongs to a discrete set $\mathcal{C}_k$; $(g)$ the RIS elements' phases are chosen between $0$ and $2\pi$; $(h)$ the CPU cycle frequency assigned to each MD is non negative; $(i)$ the sum of the CPU cycle frequencies assigned to all MDs does not exceed the maximum MEH's frequency. 

\section{Lyapunov-driven DRL: The convergence of model-based and data-driven optimization}
Due to the lack of knowledge of channels and arrivals statistics, we now apply Lyapunov stochastic optimization tools to decouple the long-term problem in (\ref{prob_form}) into a sequence of simpler per-slot problems, whose solution only requires the instantaneous observation of such context parameters. As it will be explained in the sequel, part of the per-slot problem can be solved with closed-form expressions and fast algorithms, while part of it is solved through DRL, which can efficiently handle the inherent complexity and the lack of knowledge of the involved functions. First of all, to handle constraint $(c)$, we introduce a \textit{virtual queue} $Z_k(t), \forall k\in\mathcal{K}$, whose scope is to detect instantaneous violations of the constraint, and take consequent actions to guarantee the long-term desired performance (i.e. the accuracy above a threshold). In particular, the virtual queue evolves as follows:
\begin{equation}\label{Z_evolution}
    Z_k(t+1)=\max(0,Z_k(t)-\epsilon(\mathcal{G}_k(c_k(t))-\mathcal{G}_{k,\textrm{th}})),\;\forall k\in\mathcal{K}.
\end{equation}
By definition, virtual queue $Z_k$ grows whenever the constraint is not met instantaneously, and it is drained otherwise. Also, if the mean rate stability of $Z_k$ is guaranteed, i.e. $\lim_{T\to\infty}\mathbb{E}\{Z(T)\}/T=0$, constraint $(c)$ is guaranteed \cite{Neely10}. To guarantee physical queue and virtual queue stability, we first introduce the Lyapunov Function, a scalar measure of the system's congestion state \cite{Neely10}:
\begin{equation}\label{LF}
    L((t))=\frac{1}{2}\sum_{k\in\mathcal{K}}\left[Q_k^l(t)^2+Q_k^r(t)^2+Z_k^2(t)\right],
\end{equation}
with $\mathbf{\Lambda}(t)=[\{Q_k^l(t)\}_{\forall k\in\mathcal{K}},\{Q_k^r(t)\}_{\forall k\in\mathcal{K}},\{Z_k(t)\}_{\forall k\in\mathcal{K}}]$. From \eqref{LF}, we can define the \textit{drift-plus-penalty} function, i.e. the conditional expected variation of the Lyapunov function over two successive time slots, penalized by the objective function of problem \eqref{prob_form}:
\begin{equation}\label{DPP}
    \Delta_p(\mathbf{\Lambda}(t))=\mathbb{E}\{L(\mathbf(t+1))-L((t))+V\sum_{k\in\mathcal{K}}\textrm{Tr}(\mathbf{F}_k(t))|\mathbf{\Lambda}(t)\}.
\end{equation}
Now, based on the theoretical findings in \cite{Neely10}, our method proceeds by instantaneously minimizing a suitable upper bound of \eqref{DPP} at each slot, based on instantaneous observations of time-varying context parameters, and state variables (physical and virtual queues). The proposed upper bound (whose derivations are omitted due to the lack of space), built using \cite[Eq. ($4.46$),(4.47)]{Neely10} reads as follows:
\begin{align}\label{DPP_UB}
    &\Delta_p(\mathbf{\Lambda}(t))\leq B+\mathbb{E}\bigg\{\sum_{k\in\mathcal{K}}\bigg((Q_k^r(t)-Q
    _k^l(t))\frac{\tau R_k(t)}{n_{k,b}(c_k(t))}\nonumber\\
    &-Q_k^r(t)\tau f_k(t)/w_k-Z_k(t)(\mathcal{G}_k(c_k(t))-\mathcal{G}_{k}^{\textrm{th}})\bigg)\bigg|\mathbf{\Lambda}(t)\bigg\},
\end{align}
where $B>0$ is a finite constant, whose expression is omitted due to the lack of space. Minimizing \eqref{DPP_UB} in a per-slot basis leads to two sub-problems involving communication and computation variables, respectively, whose solution is illustrated in the next paragraphs.

\subsection{Computation sub-problem}
As already mentioned in the introduction, the computation sub-problem is the same as \cite[section III.B, problem $(10)$]{Airod2022}, i.e. it does not represent a novelty in this work. The formulation involves the variables $\{f_k\}_{ k\in\mathcal{K}}$, and its solution consists on iteratively assigning computing resources to the users with the highest computation buffer load. The novelty consists in the formulation and solution of the communication sub-problem that follows.
\subsection{Communication sub-problem}
The communication sub-problem involves: i) users' covariance matrices $\mathbf{F}_k, \forall k\in\mathcal{K}$, ii) RIS reflectivity parameters $\phi_i, i=1,\ldots,M$ and iii) the compression scheme $c_k$. The problem is formulated as follows (the time index $t$ is omitted to ease the notation):
\begin{align}\label{comm_prob}
&\hspace{- .7 cm}\underset{\{\mathbf{F}_k\}_{k},\{\phi\}_{i},\{c_k\}_{k}}{\min}\;\;\mathcal{J}\\
&\textrm{subject to}\nonumber\\
&(a)\;\mathbf{F}_k\succcurlyeq 0,\;\forall k\in\mathcal{K}\quad(b)\;\textrm{Tr}(\mathbf{F}_k(t))\leq P_k,\;\forall k\in\mathcal{K}\nonumber\\
&(c)\;c_k\in\mathcal{C}_k,\;\forall k\in\mathcal{K}\quad(d)\;\; \phi_i\in[0,2\pi],\; i=1,\ldots M\nonumber
\end{align}
where $\mathcal{J}=\sum_{k\in\mathcal{K}}\bigg[(Q_k^r-Q_k^l)\frac{\tau R_k}{n_{k,b}(c_k)}+V\textrm{Tr}(\mathbf{F}_k)-Z_k\mathcal{G}_k(c_k)\bigg].$
Despite the dramatic complexity reduction with respect to the original problem \eqref{prob_form}, \eqref{comm_prob} is challenging for two main reasons: i) it is a mixed-integer non-convex program, and ii) the function $\mathcal{G}_k(c_k)$ is generally unknown. However, once $\{\phi_i\}_{\forall i}$ and $\{c_k\}_{\forall k}$ are fixed, the problem boils down to a convex problem that can be efficiently solved through a water-filling procedure that is presented in \cite{Airod2022}. Therefore, we propose to first select RIS parameters and compression schemes through a DRL-based algorithm, to get rid of the complexity introduced by discrete non-convex function and lack of model. Then, once RIS parameters and compression schemes have been set,
we solve the remaining part, which only involves the transmit covariance matrices, using a model-based approach and classical tools from convex optimization \cite{boyd2004convex}.
\par
\subsubsection{DRL-based selection of RIS reflectivity and compression}
Reinforcement Learning aims to learn a parameterized policy function (i.e. a neural network) that maps from environmental observations to available actions, so that the expected cumulative reward signal $r(t)$ is maximized.
For the communication sub-problem, the agent being trained at time $t$ observes a vector containing the system's past state variables and current channels in vectorized forms:
\begin{align}\label{eq:drl-state}
    \mathbf{s}(t) =& \Large[ \{ Q_k^l (t-1) \}, \{ Q_k^r (t-1) \}, \{ Z_k(t-1) \},  \{ \mathcal{G}_k(c_k(t-1)) \},\nonumber \\
    &  \{ f_k(t-1) \}, \{ \mathbf{H}_{k,d}(t) \}, \{ \mathbf{H}_{r,a}(t) \}, \{\mathbf{H}_{k,a}(t)\} \Large]
\end{align}
of dimensionality $5K + 2K N_a N_u + 2K M N_u 2 N_a M$ for which the real and imaginary parts of the channel vectors are received separately.
Based on $\mathbf{s}(t)$, the agent selects the action $\mathbf{a}(t) = [\{c_k(t)\}_{\forall k}, \{\phi_m(t)\}_{\forall m} ]$, which fixes the current compression and RIS profiles, allowing for the optimal values for $\{\mathbf{F}_{k}(t)\}_{\forall k}$ to be computed (see below) and consequently, for the objective value of~\eqref{comm_prob} to be calculated.
Since the RL problem is posed as maximization, we simply set $r(t) = -\mathcal{J}(t)$ to attain the equivalent optimization problem of~\eqref{comm_prob}. The system's state then proceeds to $t+1$.
\par
It is important to note that all component variables of $\mathbf{s}(t)$ appearing in~\eqref{eq:drl-state}, as well as the objective value of~\eqref{comm_prob} at time $t$ depend exclusively on the values of $\mathbf{s}(t-1)$ and $\mathbf{a}(t)$.
This remark ensures that the system evolves as a \textit{Markov Decision Process} (MDP) \cite{Pervasive}, allowing for theoretically optimal policies to be derived without the need for knowledge of its evolution from $t-2$ and backward.
Besides, the DRL formulation is algorithmic-agnostic, which allows for a wide-variety of DRL agents to be applied with different component mechanisms.
Our methodology involves the use of the well-established Proximal Policy Evaluation (PPO) algorithm \cite{PPO}, although we omit its detailed description due to space restrictions.
To make PPO applicable to the system's mechanics, we configure its policy network to output $\mathbf{a}(t)$ as vectors of continuous values in $[0,1]^{K+M}$ and we then construct $c_k(t)$ by discretizing each of the first $K$ elements to the allowed compression levels and $\phi_m(t)$ by multiplying the remaining elements by $2\pi$, as proposed in \cite{OnlineRIS}. 
\par

\vspace{-2mm}
\subsubsection{Water-filling for covariance matrix optimization}
\vspace{-1mm}

Once RIS reflectivity and compression schemes are fixed, the problem to find the optimal covariance matrix can be solved optimally. Indeed, first of all, once the RIS parameters are fixed, the problem can be decoupled across different users. Then, for all users $k\in\widetilde{\mathcal{K}}=\{\widetilde{k}\in\mathcal{K}:Q_{\widetilde{k}}^l\leq Q_{\widetilde{k}}^r\}$, the optimal solution is $\mathbf{F}=\mathbf{0}_{N_u}$, a the first two terms in \eqref{comm_prob} are non-decreasing functions of the user transmit power, as in \cite{Airod2022,Merluzzi_ICC2022}. Also, for all users $k\in\mathcal{K}\setminus\widetilde{\mathcal{K}}$, the problem is convex and can be solved through the water-filling procedure described in \cite[Algorithm 2]{Airod2022}. The overall solution seeks to balance between data rate (weighted by physical queues, i.e. delay), accuracy (weighted by virtual queues), and objective function (weighted by the trade-off parameter $V$).

\vspace{-2mm}
\section{Numerical Evaluation}
\vspace{-1mm}


\noindent\textbf{System parameters.} To illustrate the effectiveness our proposed methodology we conceive an edge inference scenario of image recognition, in the presence of an RIS, and an AP.
A ResNet32 \cite{ResNet-CVPR16} model is deployed on a MEH machine and trained to classify CIFAR-10 images, after being trained to approximately $92\%$ accuracy (over the uncompressed images), offering a reasonable compromise between performance and E2E delay.
For the shake of clarity of the results, we consider a single mobile user with no direct wireless link to the AP.
The UE performs image compression on device using the JPEG protocol, so that $c_k$ values correspond to the compression quality.
In terms of inference reliability metric, we consider the average accuracy in order to provide interpretable thresholds and performance evaluations.
While indeed, the accuracy of the network in unlabelled images cannot be known by definition, we have numerically quantified its average metric values for all possible JPEG compression levels of dataset's test set, to be used for performance evaluation. The rest of the parameter values are presented next:
Classifier parameters: $4.7 \cdot 10^5$,
$w_k=5.6$GHz,
$\bar{A_k}=4$ arrivals per time step,
$c_k \in \{1, \dots, 100\}$,
$f_{\text{max}}=3.6$ GHz,
$P_k=100$ mW,
$\tau=0.01$ s,
$W=100$ MHz,
$\mathcal{G}_{k,\textrm{th}}=0.85$,
Rice factor: $25$ dB,
$\sigma^2=-120$ dB,
operating frequency: $5$ GHz,
UE-RIS attenuation: $62.60$ dB,
AP-RIS attenuation: $66.34$ dB,
Maximum user movement displacement: $5$ m.
A frequency-division multiplexing transmission scheme is employed, where the power allocation among the narrowband frequency bins is kept equal and is not treated as part of the optimization problem. This assumption is necessary to integrate a realistic transmission paradigm without increasing the complexity of the problem.
\begin{figure}[t]
    \centering
    \includegraphics[width=0.9\linewidth]{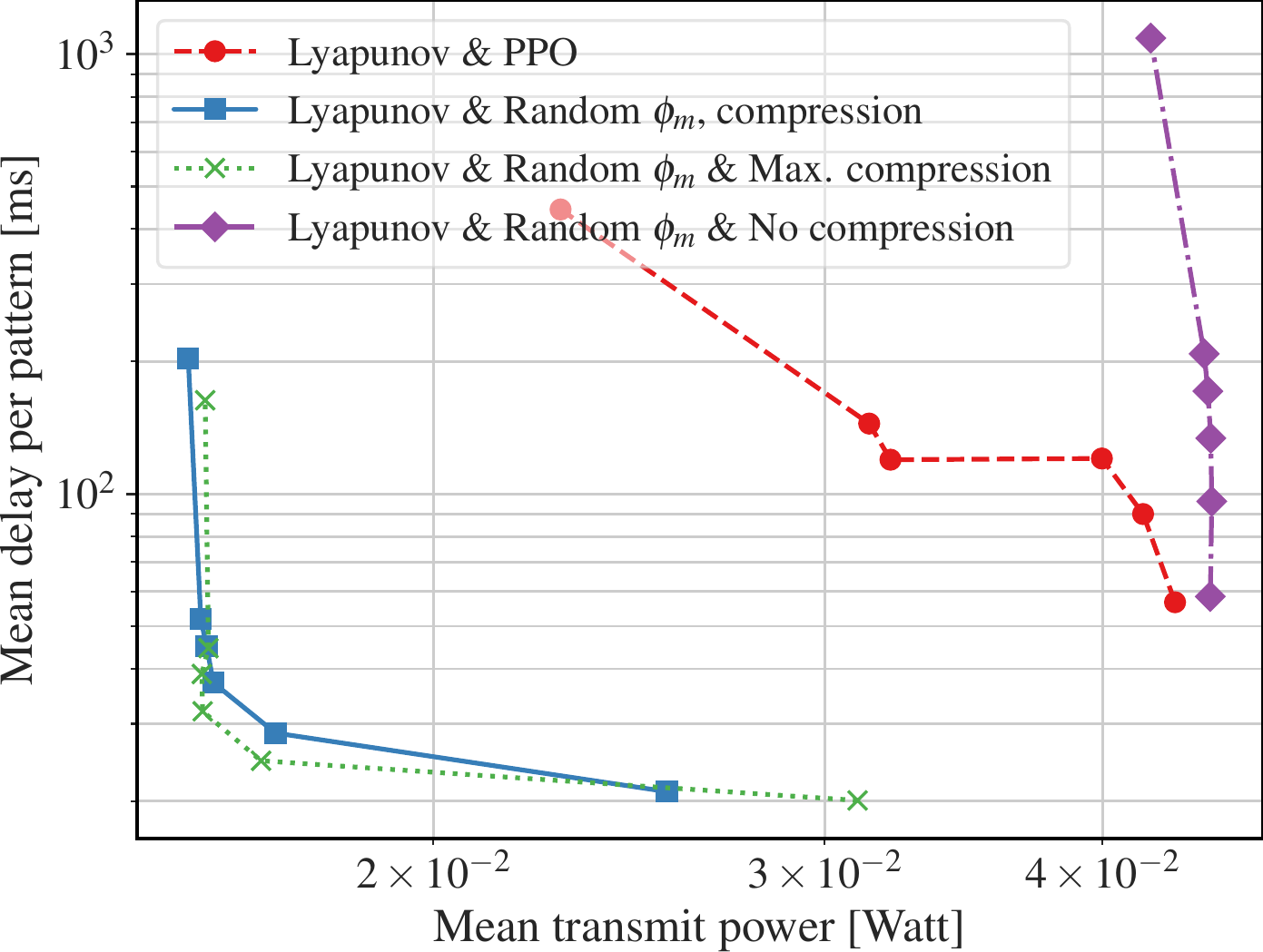}
    \caption{Evaluation of the achieved system objectives for different trade-off values of $V$. }
    \label{fig:delay_vs_power}
\end{figure}

\par
\noindent\textbf{Evaluation:} To solve the problem~\eqref{prob_form}, we jointly apply the partial methodologies presented in Section~\ref{sec:prob_formulation}. The water-filling and CPU scheduling optimization routines solve their respective sub-problems optimally at every time step.
Therefore, to assess the overall formulation we numerically evaluate performance of the DRL's selections of RIS reflections and compression schemes. As baselines, we introduce the policies of always performing (i) maximum compression (which offers the minimum delay), (ii) no compression (offering maximum accuracy) and (iii) compression at a uniformly random level, while employing the rest of the optimization components and controlling the RIS reflections at random.
Upon fixing a trade-off value for $V$ from $\{1\cdot 10^5, 2\cdot 10^5, 3\cdot 10^5, 4\cdot 10^5, 5\cdot 10^5, 3\cdot 10^6 \}$, we train a PPO instance with $5$ layers of $32$ neurons for its policy and value networks for $10^6$ time steps, resetting the system at randomized episodes of length $1500$, using the final episode as evaluation.
\par
The training phases resulted in objective values orders of magnitudes lower than the baselines, although the scales of each objective instance is heavily influenced by the choice of $V$.
The resulting accuracy offered by PPO ranges from $0.85$ to $0.91$ denoting that the DRL component always satisfies the desired constraint. The maximum and random compression schemes offer accuracy scores of $0.20$ and $0.69$, respectively.
Fig.~\ref{fig:delay_vs_power} presents the achieved delay and mean transmit power offered by each method. Compared to the no-compression approach, DRL can achieve up to $47\%$ reduction in the average UE power consumption and up to $59\%$ reduction in the maximum E2E delay.
At the same time, it results from $45\%$ to up to $2.5$ times (in the most delay-sensitive cases) higher power consumption and $5-8$ms added delay in comparison to the full compression policy, but the latter falls extremely short of any practical accuracy constraints.
Clearly, the system endowed with they Lyapunov-based techniques and DRL is able to automatically offer a balance between performance (accuracy and E2E delay) and power consumption.

\vspace{-2mm}
\section{Conclusion}
\vspace{-1mm}

In this paper, the problem of accurate, fast, and low-power edge inference has been investigated. A RIS-empowered MEC system has been proposed and a power minimization problem under E2E delay and accuracy constraints was formulated. The problem was solved through a combination of Lyapunov-driven optimization and DRL tools, showing how edge AI will play the twofold role of an efficient optimization tool and a service enabled by edge computing resources in 6G. A numerical evaluation in a RIS-empowered wireless scenario illustrated the capabilities of the methodology of achieving desired thresholds between accuracy, delay, and power consumption, striking typical the trade-off of edge AI native wireless networks. 



\bibliographystyle{IEEEbib}
\bibliography{refs}

\begin{thebibliography}{10}

\bibitem{Letaief22}
K.~B. Letaief et~al.,
\newblock ``{Edge artificial intelligence for 6G: Vision, enabling
  technologies, and applications},''
\newblock {\em IEEE J. Sel. Areas Commun.}, vol. 40, no. 1, pp. 5--36, 2022.

\bibitem{ETSIMEC}
S.~Kekki et~al.,
\newblock ``{M}{E}{C} in 5{G} networks,'' ETSI White Paper, no. 28, 2018.

\bibitem{Bi2021}
Suzhi Bi, Liang Huang, Hui Wang, and Ying-Jun~Angela Zhang,
\newblock ``Lyapunov-guided deep reinforcement learning for stable online
  computation offloading in mobile-edge computing networks,''
\newblock {\em IEEE Trans. Wireless Commun.}, vol. 20, no. 11, pp. 7519--7537,
  2021.

\bibitem{Sana2021}
M.~Sana et~al.,
\newblock ``Energy efficient edge computing: When lyapunov meets distributed
  reinforcement learning,''
\newblock in {\em Proc. IEEE ICC}, 2021, pp. 1--6.

\bibitem{Zhou19_EI}
Z.~Zhou et~al.,
\newblock ``{Edge intelligence: Paving the last mile of artificial intelligence
  with edge computing},''
\newblock {\em Proc. IEEE}, vol. 107, no. 8, pp. 1738--1762, 2019.

\bibitem{Han15}
S.~Han et~al.,
\newblock ``{Learning both weights and connections for efficient neural
  networks},''
\newblock in {\em Proc. Neural Information Processing Systems (NIPS) - Volume
  1}, Cambridge, MA, USA, 2015, NIPS'15, p. 1135–1143, MIT Press.

\bibitem{Tay18}
B.~Taylor et~al.,
\newblock ``{Adaptive deep learning model selection on embedded systems},''
\newblock {\em SIGPLAN Not.}, vol. 53, no. 6, pp. 31–43, Jun. 2018.

\bibitem{Jia18}
J.~Jiang et~al.,
\newblock ``{Chameleon: Scalable adaptation of video analytics},''
\newblock in {\em Proc. ACM SIGCOMM}, Budapest, Hungary, 2018, p. 253–266.

\bibitem{Ran18}
X.~Ran et~al.,
\newblock ``{DeepDecision: A mobile deep learning framework for edge video
  analytics},''
\newblock in {\em Proc. IEEE INFOCOM}, 2018, pp. 1421--1429.

\bibitem{Gal21}
A.~Galanopoulos et~al.,
\newblock ``{AutoML for video analytics with edge computing},''
\newblock in {\em Proc. IEEE INFOCOM}, 2021, pp. 1--10.

\bibitem{MerluzziEML2021}
M.~Merluzzi et~al.,
\newblock ``{Wireless edge machine learning: Resource allocation and
  trade-offs},''
\newblock {\em IEEE Access}, vol. 9, pp. 45377--45398, 2021.

\bibitem{Merluzzi_ICC2022}
M.~Merluzzi et~al.,
\newblock ``Energy-efficient classification at the wireless edge with
  reliability guarantees,''
\newblock in {\em Proc. IEEE ICC}, 2022, pp. 109--114.

\bibitem{Marco2019}
M.~Di~Renzo et~al.,
\newblock ``Smart radio environments empowered by reconfigurable {AI}
  meta-surfaces: {A}n idea whose time has come,''
\newblock {\em EURASIP J. Wireless Commun. Netw.}, vol. 2019, no. 1, pp. 1--20,
  2019.

\bibitem{WavePropTCCN}
G.~C. Alexandropoulos et~al.,
\newblock ``Reconfigurable intelligent surfaces and metamaterials: {T}he
  potential of wave propagation control for {6G} wireless communications,''
\newblock {\em IEEE ComSoc TCCN Newslett.}, vol. 6, no. 1, pp. 25--37, Jun.
  2020.

\bibitem{RISE-6G-EUCNC}
E.~Calvanese~Strinati et~al.,
\newblock ``Wireless environment as a service enabled by reconfigurable
  intelligent surfaces: The rise-{6G} perspective,''
\newblock in {\em Proc. EuCNC \& 6G Summit}, Porto, Portugal, 2021, pp.
  562--567.

\bibitem{RISE-6G-Commag}
E.~Calvanese~Strinati et~al.,
\newblock ``Reconfigurable, intelligent, and sustainable wireless environments
  for 6g smart connectivity,''
\newblock {\em IEEE Commun. Mag.}, vol. 59, no. 10, pp. 99--105, 2021.

\bibitem{Tsinghua_RIS_Tutorial}
M.~Jian et~al.,
\newblock ``Reconfigurable intelligent surfaces for wireless communications:
  {O}verview of hardware designs, channel models, and estimation techniques,''
\newblock {\em Intell. Converged Netw.}, vol. 3, no. 1, pp. 1--32, 2022.

\bibitem{huang2019reconfigurable}
C.~Huang et~al.,
\newblock ``Reconfigurable intelligent surfaces for energy efficiency in
  wireless communication,''
\newblock {\em IEEE Trans. Wireless Commun.}, vol. 18, no. 8, pp. 4157--4170,
  2019.

\bibitem{wu2019intelligent}
Q.~Wu and R.~Zhang,
\newblock ``Intelligent reflecting surface enhanced wireless network via joint
  active and passive beamforming,''
\newblock {\em IEEE Trans. Wireless Commun.}, vol. 18, no. 11, pp. 5394--5409,
  2019.

\bibitem{Pervasive}
G.~C. Alexandropoulos et~al.,
\newblock ``Pervasive machine learning for smart radio environments enabled by
  reconfigurable intelligent surfaces,''
\newblock {\em Proc. IEEE}, vol. 110, no. 9, pp. 1494--1525, 2022.

\bibitem{DCB}
K.~Stylianopoulos et~al.,
\newblock ``Deep contextual bandits for orchestrating multi-user {MISO} systems
  with multiple {RISs},''
\newblock in {\em Proc. IEEE ICC}, 2022, pp. 1556--1561.

\bibitem{bai2020latency}
T.~Bai et~al.,
\newblock ``Latency minimization for intelligent reflecting surface aided
  mobile edge computing,''
\newblock {\em IEEE J. Sel. Areas Commun.}, vol. 38, no. 11, pp. 2666--2682,
  2020.

\bibitem{di2021dynamic}
P.~Di~Lorenzo et~al.,
\newblock ``Dynamic edge computing empowered by reconfigurable intelligent
  surfaces,''
\newblock {\em EURASIP J. Wireless Commun. Netw.}, vol. 2022, no. 122, Dec
  2022.

\bibitem{huang2021reconfigurable}
S.~Huang et~al.,
\newblock ``Reconfigurable intelligent surface assisted mobile edge computing
  with heterogeneous learning tasks,''
\newblock {\em IEEE Trans. Cognitive Commun. Netw.}, vol. 7, no. 2, pp.
  369--382, Jun. 2021.

\bibitem{battiloro2022lyapunov}
C.~Battiloro et~al.,
\newblock ``Lyapunov-based optimization of edge resources for energy-efficient
  adaptive federated learning,''
\newblock {\em IEEE Trans. Green Commun. Netw.}, 2022.

\bibitem{Airod2022}
F.~E. Airod et~al.,
\newblock ``Reconfigurable intelligent surface aided mobile edge computing over
  intermittent mmwave links,''
\newblock in {\em Proc. IEEE SPAWC}, 2022, pp. 1--5.

\bibitem{Neely10}
M.~J. Neely,
\newblock {\em Stochastic Network Optimization with Application to
  Communication and Queuing Systems},
\newblock Morgan and Claypool, 2010.

\bibitem{boyd2004convex}
S.~Boyd and L.~Vandenberghe,
\newblock {\em Convex Optimization},
\newblock Cambridge university press, 2004.

\bibitem{PPO}
J.~Schulman et~al.,
\newblock ``Proximal policy optimization algorithms,''
\newblock {\em arXiv [cs.LG/1707.06347]}, 2017.

\bibitem{OnlineRIS}
K.~Stylianopoulos and G.~C. Alexandropoulos,
\newblock ``Online {RIS} configuration learning for arbitrary large numbers of
  1-bit phase resolution elements,''
\newblock in {\em Proc. IEEE SPAWC}, 2022, pp. 1--5.

\bibitem{ResNet-CVPR16}
K.~He et~al.,
\newblock ``Deep residual learning for image recognition,''
\newblock in {\em Conference on Computer Vision and Pattern Recognition
  (CVPR)}, Jun. 2016.

\end{thebibliography}

\end{document}